\documentclass[epj]{svjour}
\usepackage{graphics,amsmath,amssymb,multirow,color,pstricks,pst-plot}
\usepackage{xspace,graphicx,comment,hyperref,cite}

\newlength{\digitwidth} 

\newcommand{\QQbar}{\ensuremath{Q\overline{Q}}\xspace}
\newcommand{\pt}{\ensuremath{p_{\rm T}}\xspace}
\newcommand{\dd}{\ensuremath{{\rm d}}\xspace}

\newcommand{\jpsi}{\ensuremath{{\rm J}/\psi}\xspace}
\newcommand{\psip}{\ensuremath{\psi{\rm (2S)}}\xspace}

\newcommand{\myref}[1]{\ref{#1}}

\begin{document}

\title{Universal kinematic scaling\\
as a probe of factorized long-distance effects\\ 
in high-energy quarkonium production}

\titlerunning{Probing factorized long-distance effects in quarkonium production}

\authorrunning{P. Faccioli et al.}

\author{Pietro Faccioli\inst{1} 
\and Carlos Louren\c{c}o\inst{2}
\and Mariana Ara\'ujo\inst{2}
\and Jo\~ao Seixas\inst{1}
}
\institute{LIP and IST, Lisbon, Portugal
\and 
CERN, Geneva, Switzerland}

\date{Received: December 15, 2017 / Revised version: date}
\abstract{
Dimensional analysis reveals general kinematic scaling rules 
for the momentum, mass, and energy dependence of Drell--Yan and quarkonium cross sections. 
Their application to mid-rapidity LHC data provides strong experimental evidence 
supporting the validity of the factorization ansatz, 
a cornerstone of non-relativistic QCD, still lacking theoretical demonstration. 
Moreover, data-driven patterns emerge for the factorizable 
long-distance bound-state formation effects, 
including a remarkable correlation between the \mbox{S-wave} quarkonium cross sections and their binding energies. 
Assuming that this scaling can be extended to the \mbox{P-wave} case, 
we obtain precise predictions for the not yet measured feed-down fractions, 
thereby providing a complete picture of the charmonium and bottomonium feed-down structure.
This is crucial information for quantitative interpretations of quarkonium production data,
including studies of the suppression patterns measured in nucleus-nucleus collisions.
}

\maketitle


\section{Introduction}

Quarkonium production is a paradigm case study for the understanding of hadron formation, 
but the theore\-tical description of its basic mechanisms 
remains a challenge~\cite{bib:Brambilla:2010cs}.  
Both the attractiveness and the complexity of the problem 
reside in the quark-antiquark (\QQbar) binding process, 
whose degrees of freedom manifest themselves in the observed spectrum of bound states 
of different masses and spin--angular-momentum properties,
remaining unclear how they are interrelated with the 
momentum distributions measured for the different states. 
Non-relativistic QCD (NRQCD)~\cite{bib:NRQCD}, 
as well as other approaches to quarkonium production 
(e.g., colour-singlet and colour-evaporation models~\cite{bib:NRQCDvsCEM}) 
are based on a pivo\-tal postulate: the long-distance effects related to bound-state formation 
can be factorized out in the calculation of the cross sections. 
In this way, short-distance \QQbar cross sections (SDCs), 
containing the dependence on transverse momentum (\pt), rapidity ($y$), 
and pp centre-of-mass collision energy ($\sqrt{s}$), 
can be calculated perturbatively~\cite{bib:BKmodel,bib:BKMPLA,
bib:Chao:2012iv,bib:Gong:2012ug,bib:Shao:2014fca},
independently of the long-distance matrix elements (LDMEs), 
which are proportional to the probability of the transition from the 
$^{2s+1}l_j^{c}$ pre-resonance \QQbar, 
with $c=1$ (colour singlet) or 8 (colour octet), 
to the final-state observable $^{2S+1}L_J$ quarkonium.
Here, $s$, $l$, and $j$ ($S$, $L$, and $J$) denote the spin, orbital angular momentum, 
and total angular momentum of the primordial \QQbar state (quarkonium), respectively.

The LDMEs (in practice, free theory parameters) are considered constant: 
they depend on the initial \QQbar quantum numbers 
and on the bound-state properties 
(quarkonium mass $M$, quark mass $m_Q$,
velocity of the heavy quark in the quarkonium rest frame $v$, etc.), 
but not on the ``laboratory'' variables \pt, $y$, and $\sqrt{s}$, nor on the collision system. 
The factorization postulate, believed to be valid in the limit of heavy quark mass, 
for which the time of creation of the \QQbar pair, $\Delta t \sim 1/m_Q$, 
is distinctively smaller than the bound-state formation time, $\sim$\,1\,fm, 
has not yet been demonstrated from QCD principles~\cite{bib:Brambilla:2010cs}. 
Also because of inconsistencies 
between the NRQCD predictions and the measurements, 
especially concerning polarization~\cite{bib:FaccioliEPJC69,bib:FaccioliPLB736}, 
it is often argued that the validity of factorization 
might be restricted to specific kinematic domains (high \pt) 
or to the (heavier) bottomonium family.

The analysis reported in this paper addresses the problem in a data-driven way, 
comparing the patterns of momentum, mass, and collision-energy dependences 
measured for the ``reference'' case of inclusive Drell--Yan production
with the corresponding quarkonium patterns. 
It uses mid-rapidity pp LHC data and only relies on general kinematic scaling properties of the 
production cross sections.

\section{Dimensional scaling in kinematics of particle production}

The \pt and $y$ double-differential cross section 
for the inclusive production of a given particle 
in pp collisions can be written as 
\begin{equation} 
\label{eq:pp_xsect}
\frac{\dd^2 \sigma}{\dd\pt\, \dd y } = \int \frac{\dd^2\hat{\sigma}}{\dd\pt\, \dd\hat{y}}\, P(x_1)\, P(x_2)\, \dd x_1\, \dd x_2 \quad ,
\end{equation}
where $\hat{\sigma}$ is the cross section of the parton-initiated process, 
$P(x)$ are the parton density functions of the proton, 
$x_{1}$ and $x_{2}$ are the proton momentum fractions carried by the colliding partons,
and $\hat{y}$ is the particle rapidity in the parton-parton centre-of-mass frame, 
$\hat{y} = y - \frac{1}{2} \ln ({x_1}/{x_2})$. 
An implicit delta function sets the particle's \pt to the measured one. 
The cross section depends on the pp collision energy, $\sqrt{s}$, 
through the relation $x_1 x_2 = (\sqrt{\hat{s}}/\sqrt{s})^{2}$,
where $\sqrt{\hat{s}}$ is the parton-parton collision energy, function of \pt, $\hat{y}$ and $M$:
\begin{equation} 
\label{eq:sqrtshat}
\begin{split}
\sqrt{ \hat{s} }/M & = \hat{E}/M + \hat{p}/M \quad , \\
\hat{E}/M & = \sqrt{1 + ({\pt}/{M})^2} \, \cosh\hat{y} \quad , \\
\hat{p}/M & = \sqrt{\sinh^2\hat{y} + ({\pt}/{M})^2\cosh^2\hat{y}} \quad .
\end{split}
\end{equation}

We start by discussing inclusive Drell--Yan lepton-pair production. 
After integrating over the lepton emission angles 
in the dilepton rest frame 
(and over the degrees of freedom of the recoil system), 
the kinematics in the rest frame of the colliding partons is fully described by \pt, $\hat{y}$, and $M$.
Since the dimensionality of the differential partonic cross section is $[\mathrm{energy}]^{-3}$,
we can write
\begin{equation} 
\label{eq:xsectFormulaDYpartonic}
\frac{\dd^2\hat{\sigma}}{\dd\pt\, \dd\hat{y}} = \frac{1}{M^3} \; f(\xi,\hat{y}) \quad ,
\end{equation}
$f$ being a dimensionless function of the dimensionless variables $\hat{y}$ and $\xi$,
where we use the definition $\xi \equiv \pt/M$ to have more compact equations.
When the mixture of production mechanisms does not change with $M$, 
the partonic cross section, 
calculated at any arbitrary reference kinematic point $(\xi^\star,\hat{y}^\star)$, 
scales as $M^{-3}$. 
This property can be tested using Drell--Yan cross sections, measured
in a range of masses sufficiently smaller than the $Z$ boson mass
to stay unaffected by the strongly mass-dependent interference 
of $\gamma$- and $Z$-exchange mechanisms.
The $M$ scaling of the observable cross section 
depends on the pp collision energy via the parton densities, 
an effect that can be 
singled out in a purely data-driven way 
at LHC energies and mid rapidity, given the 
small average $x_{1,2}$.

For illustration, we parametrize $P(x)$ as
\begin{equation}
P(x) = A \cdot x^{-B} \cdot q(x) \quad ,
\end{equation}
explicitly showing the factor describing the low-$x$ behaviour and 
globally representing the remaining factors by the function $q$,
typical examples of which being
$q(x) = (1-x)^C\, (1 + D \sqrt{x} + E x)$ or 
$q(x) = (1-x)^C\, (1 + D x^E)$, 
all tending to unity in the small-$x$ limit.
Therefore,
\begin{equation} 
\label{eq:P1P2}
\begin{split}
P(x_1)\, & P(x_2) = A^2 \frac{1}{(x_1 \, x_2)^B}\, q(x_1)\, q(x_2) \\
& = A^2 \left(\frac{\sqrt{s}}{M}\right)^{2B} \left(\frac{\sqrt{\hat{s}}}{M}\right)^{-2B} q(x_1)\, q(x_2) \quad .
\end{split}
\end{equation}
Moreover,
\begin{equation}
\dd x_1 \dd x_2 = 2 \left(\frac{\sqrt{s}}{M}\right)^{-2} 
\left(\frac{\sqrt{\hat{s}}}{M}\right)^{2}  
\frac{\hat{E}/M}{\hat{p}/M} \, 
\frac{\xi}{1+\xi^2} \, 
\dd\xi \, \dd{\hat{y}_0} ~ ,
\end{equation}
%
%
%
%
where $\hat{y}_0 = y - \hat{y} = \frac{1}{2} \ln ({x_1}/{x_2})$ is the rapidity of the system of colliding partons. 
Using the previous relations and releasing the $\xi$ integration in Eq.~\myref{eq:pp_xsect}, 
we obtain the pp (``dressed'') version of the partonic cross section, Eq.~\myref{eq:xsectFormulaDYpartonic},
\begin{equation} \label{eq:xsectFormulaDY}
\frac{\dd^2 \sigma}{ \dd\pt \,\dd y } = 
\frac{1}{M^3} \left(\frac{\sqrt{s}}{M}\right)^{b} 
F\left(\xi, y; \frac{\sqrt{s}}{M} \right) \quad ,
\end{equation}
where $b = 2B - 2$ and
\begin{equation}
F\left(\xi,  y; \frac{\sqrt{s}}{M} \right) =
\int g\left(\xi, \, y-\hat{y}_0\right) q(x_1)\, q(x_2) \, \dd \hat{y}_0 \quad .
\end{equation}
The function $g$ absorbs
all the factors depending on $\xi$ and $\hat{y}$, 
including those contained in $\sqrt{ \hat{s} }/M$, $\hat{E}/M$, and $\hat{p}/M$ 
(Eq.~\myref{eq:sqrtshat}).
$F$ is a fully experimentally determinable dimensionless function 
of the dimensionless scaling variables $\xi$ and $y$. 
It also depends on $\sqrt{s}$ and $M$, but always through the $\sqrt{s}/M$ ratio, via the terms 
$q(x_{1,2}) = q( \sqrt{\hat{s}}/\sqrt{s} \,\, e^{\pm \hat{y}_0})$ 
and the integration extremes 
$\pm \ln(\sqrt{s}/\sqrt{\hat{s}_{\rm min}}\,) = 
\pm \ln(\sqrt{s}/M) \times {(1 - {\ln(\xi+\sqrt{1+\xi^2})} \,/\, {\ln(\sqrt{s}/M)} )}$. 
The sensitivity of $F$ on $\sqrt{s}/M$ decreases with increasing $\sqrt{s}$ and towards mid rapidity.
We normalize $F$ over the $(\xi, y)$ range of the analysis and 
assimilate its integral $I(\sqrt{s}/M)$ in a numerical redefinition of the $b$ exponent
(always possible to a very good approximation over a small range of $\sqrt{s}$ values, 
as in our analysis of $\sqrt{s}=7$ and 8\,TeV data),
\begin{equation} \label{eq:xsectFormulaDY_final}
\frac{\dd^2 \sigma}{ \dd\pt \,\dd y } = 
\frac{1}{M^3} \left(\frac{\sqrt{s}}{M}\right)^{\beta} 
{\cal F}\left(\xi, y; \frac{\sqrt{s}}{M} \right) \quad ,
\end{equation}
with $\int {\cal F} \,\dd\xi \, \dd y = 1$. 
The $(\pt/M, y)$-integrated cross section becomes $\sigma = M^{-2} \, (\sqrt{s}/M)^{\beta}$ 
and its measurement at two or more $\sqrt{s}$ values provides a determination of $\beta$. 

These considerations lead to the following \textit{proposition}.
In a domain where Drell--Yan production mechanisms do not interfere in varying proportions, 
the joint distribution of the 
variables $\pt/M$, $y$, and $\sqrt{s}/M$ is \textit{universal}: 
its shape does not depend on $M$.
At any given kinematic point $((\pt/M)^\star,y^\star)$ and $\sqrt{s}$, 
the $\dd \sigma / \dd\pt$ cross section scales like $M^{-(3+\beta)}$, 
where $\beta$ expresses the increase of the integrated cross section with $\sqrt{s}$.
The scaling of the cross section with $M$, integrated over \pt, 
was previously discussed in Ref.~\cite{bib:Craigie},
but using process-specific arguments rather than completely general dimensional-analysis considerations.

\begin{figure}[t]
\begin{center}
\includegraphics[width=0.87\linewidth]{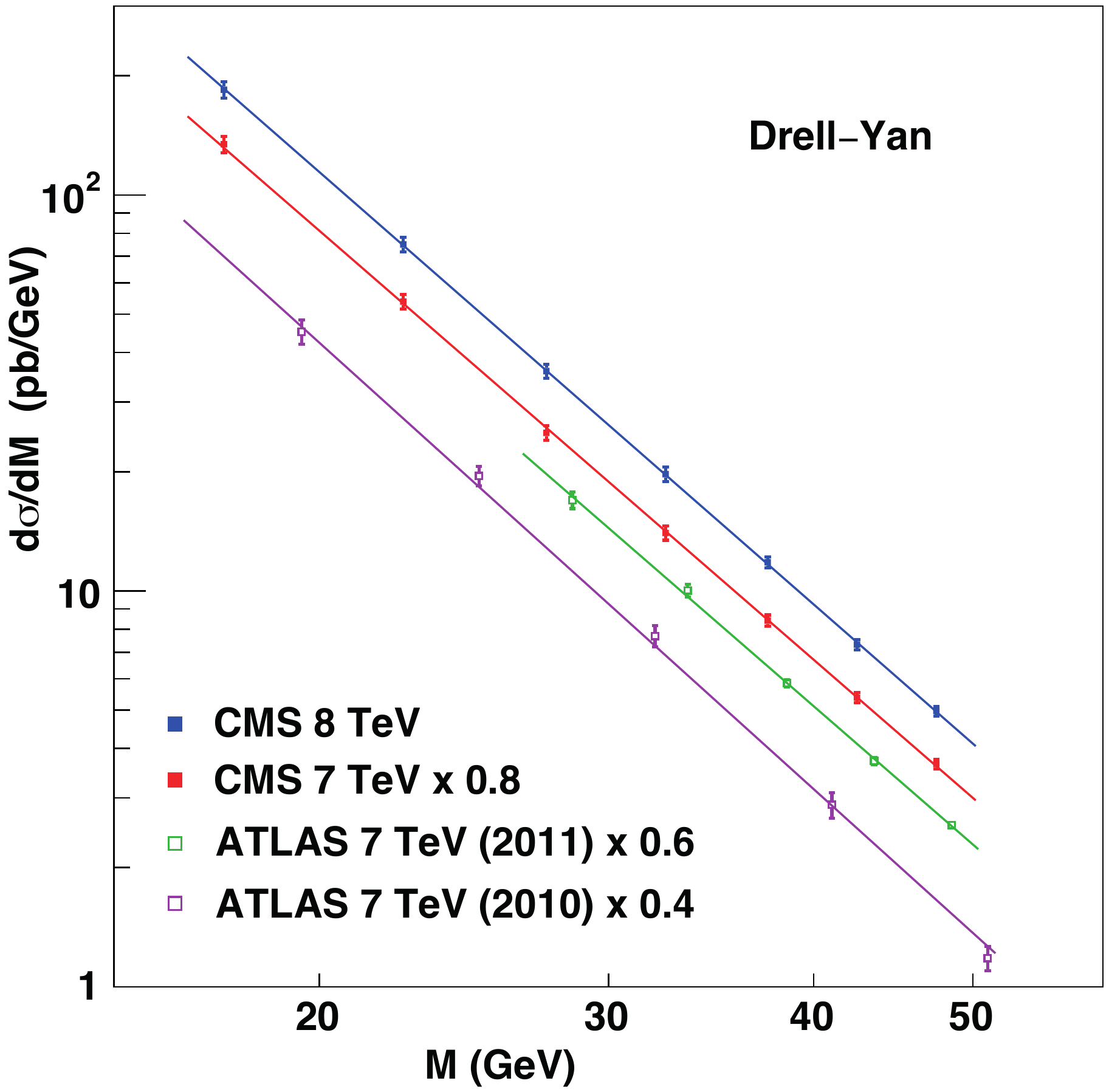}
\caption{Mid-rapidity Drell--Yan differential cross sections as a function of the dilepton mass,
as measured by ATLAS~\cite{bib:DY_ATLAS} and CMS~\cite{bib:DY_CMS7,bib:DY_CMS8}.}
\label{fig:DY}
\end{center}
\end{figure}

We will now consider the Drell--Yan $\dd\sigma/\dd M$ differential cross sections 
measured by ATLAS~\cite{bib:DY_ATLAS} and CMS~\cite{bib:DY_CMS7,bib:DY_CMS8},
in pp collisions at $\sqrt{s}=7$ and 8\,TeV, 
shown in Fig.~\myref{fig:DY} for the $M < 50$\,GeV range. 
It can be recognized from the previous discussion 
that $\dd\sigma/\dd M$ has the same $M^{-3}\,(\sqrt{s}/M)^{\beta}$ 
scaling behaviour as $\dd\sigma/\dd\pt$ at fixed $\pt/M$ and $y$.
Simultaneously fitting all four data sets to a single $M^{-\alpha_{\mathrm{DY}}}$ function, 
only considering point-to-point uncorrelated uncertainties
in order to determine the shape parameter with maximum significance, 
gives a remarkably precise result: $\alpha_{\mathrm{DY}} = 3.63 \pm 0.03$. 
Fitting each data set individually gives: 
$3.60 \pm 0.05$ (CMS 7\,TeV), 
$3.63 \pm 0.05$ (CMS 8\,TeV), 
$3.75 \pm 0.07$ (ATLAS 7\,TeV, 2010) and 
$3.60 \pm 0.07$ (ATLAS 7\,TeV, 2011).

The $\beta$ exponent, reflecting the $\sqrt{s}$ dependence, can be derived from the ratio between 
the cross sections measured at 8 and 7\,TeV, directly reported by CMS~\cite{bib:DY_CMS8},
\mbox{leading} to $\beta = 0.73 \pm 0.15$, 
where the uncertainty is obtained assuming a relative uncorrelated luminosity uncertainty of 2\%. 
The resulting $\alpha-\beta$ difference, $2.90 \pm 0.15$, 
is perfectly consistent with 3, as expected.

The Drell--Yan cross sections so far reported by the LHC experiments are integrated over \pt.
The future availability of \pt- and $\pt/M$-differential measurements will allow the realisation of 
more detailed and accurate tests.
It is worth noting that the existence of these simple and general kinematic scaling rules
has been ignored in all the published analyses of LHC data, 
an unfortunate situation because of the resulting loss in physics value and 
also because they can be very useful experimental tools.
In fact, detector and trigger acceptances drastically sculpt the reconstructed dilepton distributions,
especially in the low momentum and low mass regions. 
Verifying that the measured $\pt/M$ and $M$ spectra, after efficiency corrections, etc.,
satisfy the expected dimensional scaling 
constitutes a powerful cross-check of the analysis procedure
and a validation of the reported systematic uncertainties.

\section{Dimensional scaling in quarkonium production}

Moving now to our main case study, quarkonium production,
we write the differential cross section for the inclusive production of a given \mbox{S-wave} state as
\begin{equation} 
\label{eq:xsectQuarkoniumPartonicGeneral}
\begin{split}
\frac{\dd^2 \sigma}{\dd \pt\, \dd y} &
= m_Q^{-3}  \left(\frac{\sqrt{s}}{M}\right)^{\beta} 
\sum_i \frac{\mathcal{L}_i(m_Q, M, \xi, y, \sqrt{s}/M)}{m_Q^3} \\
 & \times \; \mathcal{F}_i (m_Q, M, \xi, y, \sqrt{s}/M) ) \; ,
\end{split}
\end{equation}
where the overall factor $m_Q^{-3}$ matches the global dimensionality of the observable. 
The only specification of the functions $\mathcal{L}_i$ is that they have dimension $[\mathrm{energy}]^3$, 
formally compensated by the $m_Q^3$ denominators, 
while the $\mathcal{F}_i$ are dimensionless shape functions (defined with $\int {\cal F} \,\dd\xi \, \dd y = 1$, as in the Drell--Yan case). 
The $\sqrt{s}/M$ power-law factor represents the modification from partonic to observable level, 
as from Eq.~\myref{eq:xsectFormulaDYpartonic} to Eq.~\myref{eq:xsectFormulaDY_final}. 
The coefficient $\beta$ is the same as for Drell--Yan production. 
From the previous discussion, we can obtain a precise evaluation of its value at $\sqrt{s} \simeq 7$ or 8\,TeV as the difference between the 
average experimental mass scaling exponent $\alpha_{\mathrm{DY}}$ and the expected bare-cross-section scaling exponent: 
$\beta = \alpha_{\mathrm{DY}} - 3 = 0.63 \pm 0.03$.

The expression above, Eq.~\myref{eq:xsectQuarkoniumPartonicGeneral}, 
is built in analogy to the NRQCD factorized expansion.
In this limit, $\mathcal{F}_i$ and $\mathcal{L}_i$ encode the information on, respectively, 
the hard scattering process producing a \QQbar pair, of mass $\simeq 2\, m_Q$, 
and its ensuing long-distance transition to the final bound state $\mathcal{Q}$, of mass $M$. 
In NRQCD the LDMEs $\mathcal{L}_i$ are independent of laboratory kinematics 
and only depend on degrees of freedom naturally defined in the $\mathcal{Q}$ rest frame 
($M$, $m_Q$, $v$, quantum numbers, spectroscopic energy levels, etc.), 
while the kinematic dependence is contained in the SDCs, corresponding to $\mathcal{F}_i / m_Q^6 (\sqrt{s}/M)^{\beta}$, 
where the index $i$ indicates one specific 
$\QQbar(^{2s+1}l_j^{c}) \to \mathcal{Q}(^{2S+1}L_J)$ production channel. 

In this formulation, a consequence of the factorization hypothesis is that 
each $\mathcal{F}_i$ function should be independent of $M$, being,
in particular, the same for all directly produced $^3S_1$ charmonia and bottomonia. In fact, it is important to note that
the mass difference between the $\mathcal{Q}$ and \QQbar states plays no role in the $\pt/M, \hat{y}$ dependence of $\mathcal{F}_i$, 
given that, similarly to the physical transitions between states of the same quarkonium family~\cite{bib:FaccioliPLB773}, 
the $\QQbar \to \mathcal{Q}$ transition preserves the ratio between laboratory-vector-momentum and mass 
(and, therefore, both $\pt/M$ and $\hat{y}$).

\begin{figure*}[!t]
\begin{center}
\includegraphics[width=0.64\linewidth]{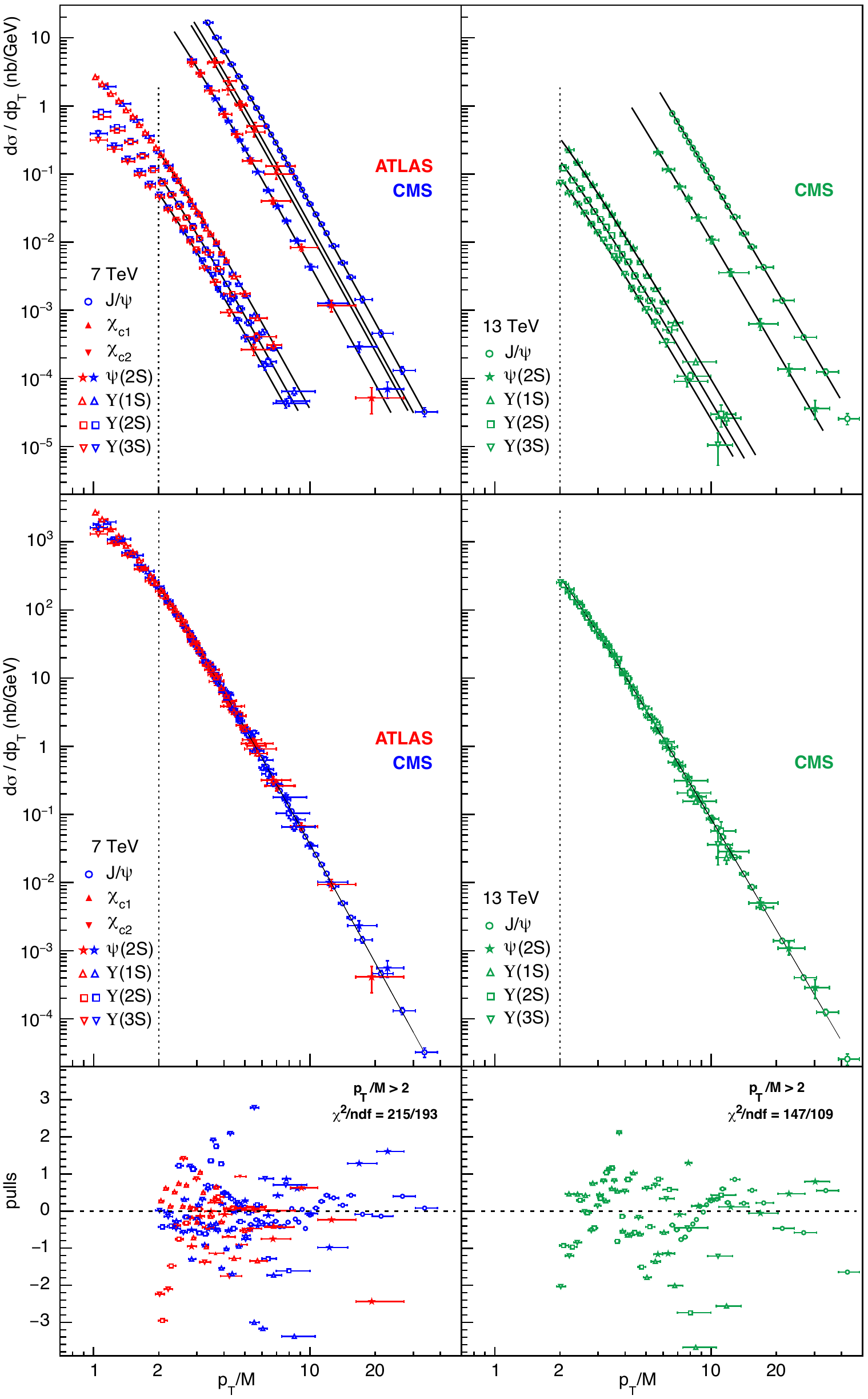}
\caption{Mid-rapidity \emph{prompt} quarkonium cross sections in pp collisions
at $\sqrt{s} = 7$\,TeV (left) and 13\,TeV (right),
as measured by ATLAS (red markers)~\cite{bib:ATLASpsi2S, bib:ATLASYnS, bib:ATLASchic}
and CMS (blue and green markers)~\cite{bib:CMSjpsi, bib:CMSYnS, bib:BPH15005} (top panels),
and after scaling up all the normalizations to match those of the \jpsi\ cases (middle panels).
For each collision energy, the curves represent a single universal empirical function, 
of shape determined by a simultaneous fit to all data points of $\pt/M > 2$
and normalizations specific to each quarkonium state.
The bottom panels show the pulls between each data point and the fitted function.}
\label{fig:pTovMscaling}
\end{center}
\end{figure*}

While these notations have been chosen to accommodate the NRQCD factorization expansion as a limit case, 
Eq.~\myref{eq:xsectQuarkoniumPartonicGeneral}, with generic $\mathcal{L}_i$ and $\mathcal{F}_i$ 
both redundantly depending of the relevant variables, 
represents a fully general template of the cross section, with no prejudice on its physical scaling properties.

Very interesting physical indications transpire from the unforeseen simplicity of the trends measured by ATLAS and CMS,
as discussed in the next paragraphs.

%
%

The first experimental input to our considerations are the quarkonium cross sections measured 
at mid-rapidity ($|y| \lesssim 2$) by the ATLAS and CMS experiments, shown in the top panels of
Fig.~\myref{fig:pTovMscaling} for $\sqrt{s} = 7$\,TeV (left) and 13\,TeV (right).
All the measurements, from the \jpsi\ to the $\Upsilon$(3S), 
scale with $\pt/M$ in a state-independent way, at least for not very low \pt ($\pt/M \gtrsim 2$). 
This ``universality'' is well illustrated by the middle panels, which clearly show that,
after rescaling the state-specific normalizations to those of the \jpsi,
the cross section shapes become indistinguishable from each other.
The goodness of this universal scaling can be better appreciated by looking at the bottom panels,
which show (in a linear scale) the pull distributions, 
i.e.\ the differences between each data point and the universal fitted function, 
divided by the measurement uncertainty.
No systematic trends are seen in the pull distributions and the observed deviations are
very well compatible with statistical fluctuations.
The cross section fits consider correlations between the luminosity uncertainties in each data set 
and the pulls, evaluated to check the consistency with one common shape, 
are calculated excluding such uncertainties.

\begin{figure}[t]
\begin{center}
\includegraphics[width=0.9\linewidth]{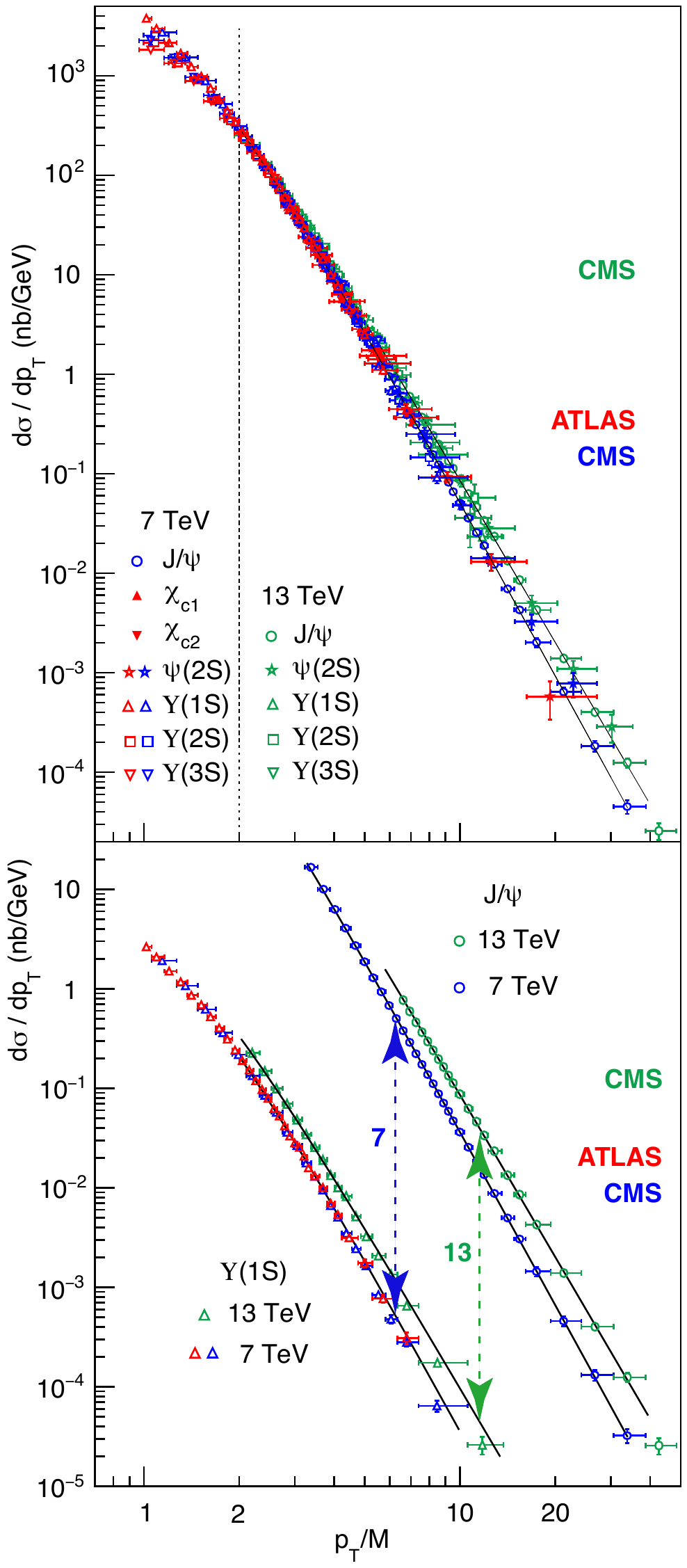}
\caption{Shape comparison between the 7 and 13\,TeV $\pt/M$-differential quarkonium cross sections (top)
and normalization comparison between the \jpsi\ and $\Upsilon$(1S) at the two energies (bottom).}
\label{fig:energy_and_mass_dependence}
\end{center}
\end{figure}

A slightly broader $\pt/M$ distribution is observed at the higher energy, 
as shown in Fig.~\myref{fig:energy_and_mass_dependence}-top.
The fact that the distinct P- and \mbox{S-wave} states 
show a compatible kinematic scaling is discussed in Ref.~\cite{bib:chic_polarization_difference}.
Here we focus on the precisely measured cross sections of the closely related $^3S_1$ states. 
Their universal $\pt/M$ scaling, at a given energy, indicates that the same mixture of processes 
(or one common dominating process) describes the production of all these states.

The second experimental indication for our following discussion is the mass scaling of the cross section from charmonium to bottomonium.
By exploiting the $\pt/M$ scaling we can determine it without
relying on model-dependent extrapolations to low \pt:
it is sufficient to consider the fitted $\pt/M$ distributions at an arbitrary $\pt/M = (\pt/M)^\star$. 
We consider for each family the meson closest to the ground state, \jpsi\ and $\Upsilon$(1S), 
at the two energies (Fig.~\myref{fig:energy_and_mass_dependence}-bottom).

To obtain the yield of \emph{directly} produced \jpsi\ mesons, 
we subtract from the fitted prompt-\jpsi\ normalization 
those of the $\chi_{c1}$, $\chi_{c2}$ and \psip\ states 
(always considered at the same $\pt/M$ value: as mentioned before, $\pt/M$ is transferred unchanged from mother to daughter), 
weighted by the branching fractions~\cite{bib:PDG} of the respective feed-down processes. 
The result is a \jpsi\ direct-production fraction of 31.9\%.
For the $\Upsilon$(1S) we assume a direct-production fraction of $(50 \pm 10)\%$. 
Later in this paper we will need the corresponding fractions for the $\Upsilon$(2S) and $\Upsilon$(3S),
which we assume to be 60\% and 70\%, respectively, also with $\pm 10$\% uncertainties.
These values were estimated from $\Upsilon$(nS) yield ratios measured by ATLAS~\cite{bib:ATLASYnS} and CMS~\cite{bib:CMSYnS},
complemented by LHCb data on the (large) $\chi_b{\rm (mP)} \to \Upsilon{\rm (nS)}$ feed-down fractions~\cite{bib:LHCbChibFeedown}.

The resulting mass scaling, measured as
\begin{equation}
\frac{\left. \dd \sigma / \dd \pt (\Upsilon(\mathrm{1S}))\right._{\xi = \xi^\star} }    
{\left. \dd \sigma / \dd \pt (\mathrm{J}/\psi)\right._{\xi = \xi^\star} } = \left(\frac{m_b}{m_c}\right)^{-\alpha} \; , 
\label{eq:massDependenceUpsilonVsJpsi}
\end{equation} 
does not show a significant energy dependence: 
$\alpha = 6.6 \pm 0.1$ and $6.5 \pm 0.1$ at 7 and 13\,TeV, respectively,
with $2\, m_c \simeq M(\eta_c{\rm (1S)}) = 2.984$~GeV and $2\, m_b \simeq M(\eta_b{\rm (1S)}) = 9.389$~GeV~\cite{bib:PDG}. 
Subtracting the $\beta$ value mentioned above, $0.63\pm0.03$, measured in Drell--Yan production at 7 and 8\,TeV, 
we find that the partonic-level differential cross section changes as $\simeq m_Q^{-6}$ 
between the charmonium and bottomonium families. 

These experimental facts constrain and specify the functions $\mathcal{L}$ and $\mathcal{F}$ 
appearing in Eq.~\myref{eq:xsectQuarkoniumPartonicGeneral}.
The independence of the $\pt/M$ scaling on either $M$ or $m_Q$, at a given energy, 
indicates that the $\pt/M$ and $(M,m_Q)$ dependences do not ``mix'': 
$\mathcal{L} \; \times \; \mathcal{F}$ = $\mathcal{L}(m_Q, M, \sqrt{s}/M) \; \times \; \mathcal{F} (\xi, y, \sqrt{s}/M )$ 
(further studies considering forward-rapidity, low-\pt data will assess the combined $\pt/M, y$ dependence of $\mathcal{F}$, 
here effectively a function of only $\pt/M$). 
In other words, there is experimental evidence that, at mid rapidity and not too low \pt, 
it is possible to describe quarkonium production ``factorizing'' short- and long-distance effects, 
\textit{defined} as the dependences on, respectively, the laboratory momentum of the detected meson ($\mathcal{F}$) 
and the specific bound-state properties ($\mathcal{L}$).

The measured $m_Q^{-6}$ scaling of the differential cross section at a fixed $\pt/M$ (and $y$) further specifies $\mathcal{L}$.
By precisely equating the explicit $m_Q$ dependence of the expression in Eq.~\myref{eq:xsectQuarkoniumPartonicGeneral} 
(coming from the overall factor and the denominators of the $\mathcal{L}$ terms, with $\mathcal{F}$ now taken independent of $m_Q$), 
such result leaves no margin for a dependence of $\mathcal{L}$ on $m_Q$, 
if not counterbalanced by a dependence on $M$ and/or $\sqrt{s}$: 
$\mathcal{L} = \mathcal{L}(m_Q / M, m_Q / \sqrt{s})$. 
However, given that no significant difference in mass scaling is observed at $7$ and $13$\,TeV, 
a $m_Q / \sqrt{s}$ dependence would be experimentally indistinguishable 
from the analogous global energy scaling of all quarkonium cross sections already accounted for 
by the $\beta$ power law in Eq.~\myref{eq:xsectQuarkoniumPartonicGeneral}.
It is therefore always possible to \emph{define} the long-distance factors as $\sqrt{s}$-independent, 
thereby reducing them to functions of the kind $\mathcal{L} = \mathcal{L}(m_Q / M)$, 
as is illustrated by the discussion starting in the next paragraph. 

It should be clear from these considerations that such $\mathcal{L}$ functions do not coincide with the LDMEs of NRQCD: 
instead of being defined by setting an energy scale within the theory, 
they are built on the basis of dimensional-analysis and data patterns, 
becoming universal, experimentally definable quantities. 
Apart from differences in operative definitions, 
we note that the remarkably simple picture of \mbox{S-wave} quarkonium production emerging from the data 
mirrors well the primary concepts of NRQCD factorization and of universality of the long-distance bound-state formation effects. 
A future verification of the corresponding experimental patterns in different collision systems can further probe these fundamental concepts. 

By adopting now the guiding idea of factorization of short- and long-distance effects, 
as supported by data, we can analyse in more detail the mass scaling, using all \mbox{S-wave} states. 

\begin{figure}[t]
\begin{center}
\includegraphics[width=0.87\linewidth]{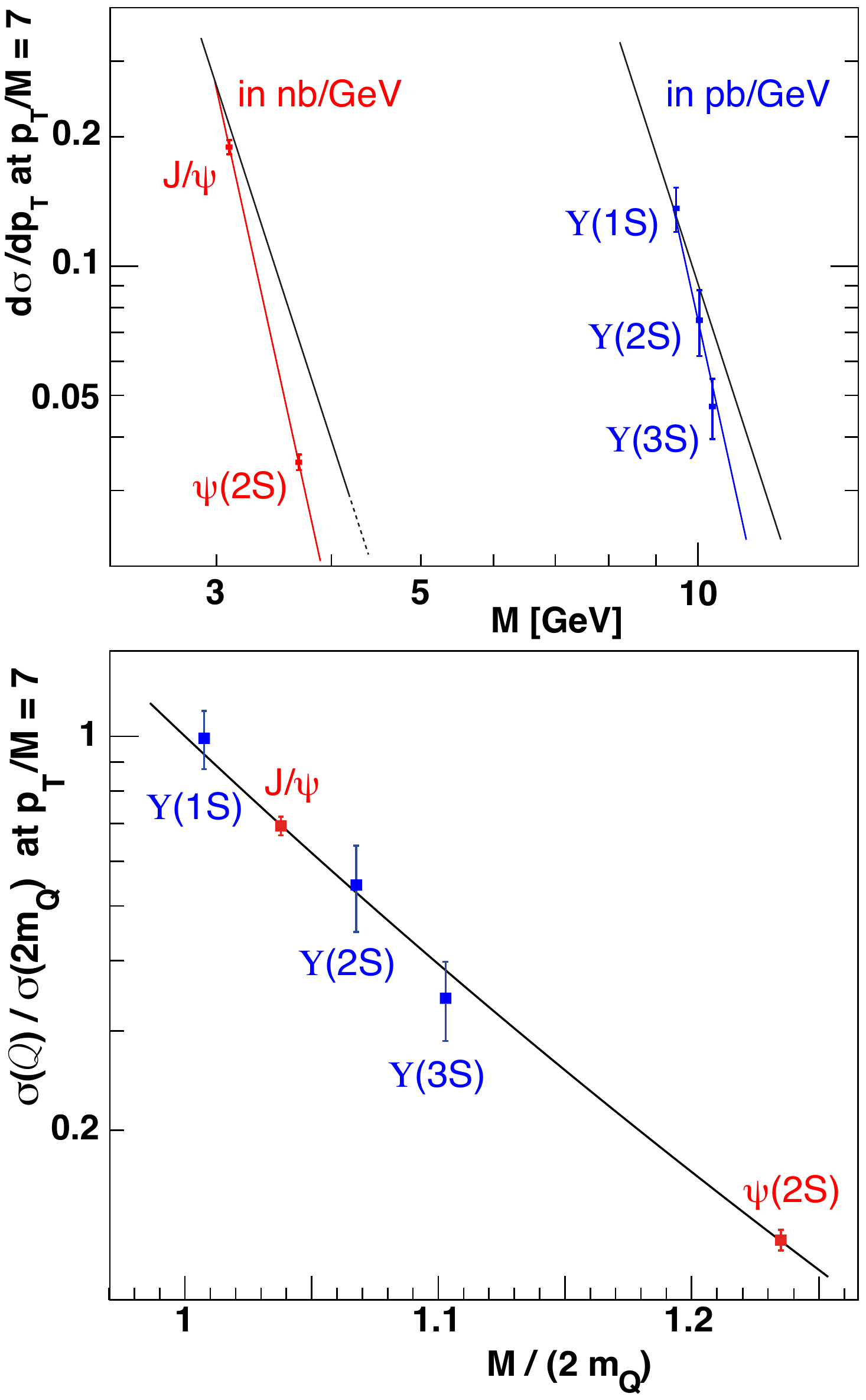}
\caption{Mid-rapidity \emph{direct} production cross sections of $^3S_1$ quarkonia, 
at $\sqrt{s} = 7$\,TeV and $\pt/M = 7$.
In the top panel, the (single) black line is the result of a fit to the five cross sections, 
after extrapolating them (red and blue lines) to $\sigma(2\,m_Q)$. 
The bottom panel shows the common trend of the cross sections, normalized to the extrapolated values, 
for the five quarkonium states.}
\label{fig:Mdependence}
\end{center}
\end{figure}

Figure~\myref{fig:Mdependence}-top shows the values of the \emph{direct} production cross sections 
of the \mbox{S-wave} quarkonia, for one specific and arbitrarily chosen point, $\pt/M = 7$, 
as derived from the ${\cal F}(\pt/M)$ functions shown in the top panel of Fig.~\myref{fig:pTovMscaling}.
Given the universality of the function, a different $\pt/M$ reference point only leads to an overall vertical rescaling
of the points in Fig.~\myref{fig:Mdependence}, not affecting the mass dependence itself.
For the \jpsi\ and $\Upsilon$(nS), the \emph{prompt} cross sections were scaled down
by the direct-production fractions mentioned earlier.
%
%
The figure only shows the $7$\,TeV pattern; 
completely analogous results are found in the 13\,TeV data (apart from a significantly different global normalization scale), 
as discussed hereafter.

From the \jpsi\ to the $\Upsilon$(3S) we observe, 
besides the global drop by three orders of magnitude, 
a ``fine structure'': 
within each quarkonium family, the cross section decreases faster than from one family to the other.
In the factorization perspective, the global drop represents the short-distance scaling, 
reflecting the change from $M(\QQbar) \simeq 2\, m_c$ to $\simeq$\,$2\, m_b$.
The steeper change within each family reflects the $M$-dependent $\QQbar \to \mathcal{Q}$ transition probability 
$\mathcal{L}(m_Q, M)$. 
The factorized interpretation is represented by the fit curves shown in the figure. 
First (red and blue curves) the charmonium and bottomonium cross sections are extrapolated, 
respectively, to $2\, m_c$ and $2\, m_b$ (as defined above).
From the resulting $\sigma(2m_b) \,/\, \sigma(2m_c)$ ratio, 
the short-distance charmonium-to-bottomonium mass scaling (black curve) is determined as $\propto m_Q^{-\alpha_{\QQbar}}$, 
with $\alpha_{\QQbar} = 6.63 \pm 0.08$, practically identical to the previously obtained result, 
considering only the \jpsi\ and the $\Upsilon$(1S), leading to the $m_Q^{-6}$ partonic-level scaling discussed above.

Here we focus on the long-distance mass dependence. 
The bottom panel of Fig.~\myref{fig:Mdependence} shows 
the dependence of the $\sigma(\mathcal{Q})/\sigma(2\,m_Q)$ ratio on $M/(2\,m_Q)$: 
one common power-law, $\sigma(\mathcal{Q})/\sigma(2\,m_Q) = [M/(2\,m_Q)]^{-(9.7 \pm 0.3)}$, 
describes very well both the $\psi$ and the $\Upsilon$ data points.

\begin{figure}[t]
\begin{center}
\includegraphics[width=0.98\linewidth]{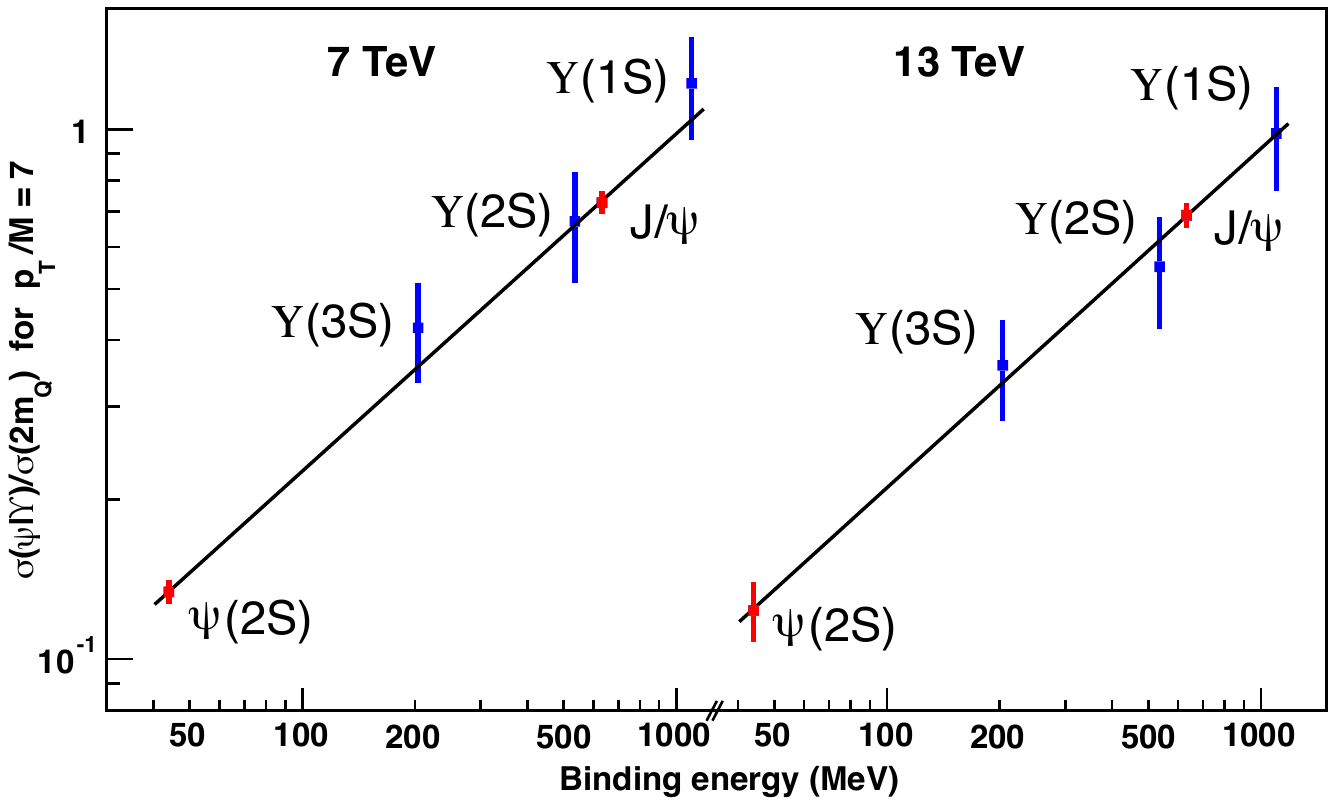}
\caption{Correlation between the long-distance factor $\sigma(\mathcal{Q})/\sigma(2\,m_Q)$ 
and the quarkonium binding energy, at 7 and 13\,TeV.}
\label{fig:crosssectionratio}
\end{center}
\end{figure}

Interestingly, as shown in Fig.~\myref{fig:crosssectionratio}, 
this monotonic dependence can be seen as a tight correlation of the long-distance factors to the quarkonium binding energy, 
defined as $2M(D^0) - M(\psi{\rm (nS)})$ or $2M(B^0) - M(\Upsilon{\rm (nS)})$ 
and calculated with mass values from Ref.~\cite{bib:PDG}.
As a function of this variable we see another ``universal'' trend: 
the data points are well described by the parametrization $\sigma(\mathcal{Q})/\sigma(2\,m_Q) \propto E_{\mathrm{binding}}^{\delta}$, 
common to charmonium and bottomonium and 
identical at 7\,TeV ($\delta = 0.63 \pm 0.02$) and 13\,TeV ($\delta = 0.63 \pm 0.04$). 
The central values of the $\delta$ and $\beta$ parameters are identical by coincidence.
This result gives further support to the idea that the dependence of the cross sections on the bound-state masses 
is a ``factorizable" long-distance effect, independent of the laboratory-momentum.

\begin{table}[!htp]
\centering
\caption{Feed-down fractions between the several quarkonia,
for $\pt/M > 2$.
The first uncertainty reflects the precision of our fit, 
while the second reflects the branching fractions.}
\label{tab:feeddowns}
\begin{tabular}{ccc}
\hline\noalign{\vglue0.85mm}
Daughter & Parent & Feed-down fraction (\%)\\
\noalign{\vglue0.85mm}\hline\noalign{\vglue0.85mm}
\jpsi          & $\chi_{c0}$   &   0.762 $\pm$ 0.046 $\pm$ 0.036 \\
                & $\chi_{c1}$   &  15.61 $\pm$ 0.87 $\pm$ 0.47 \\
                & $\chi_{c2}$   &   7.83 $\pm$ 0.46 $\pm$ 0.27 \\
                & \psip   &   7.67 $\pm$ 0.88 $\pm$ 0.08 \\
\noalign{\vglue0.85mm}\hline\noalign{\vglue0.85mm}
$\chi_{c0}$ & \psip  &    2.09 $\pm$ 0.25 $\pm$ 0.06 \\
\noalign{\vglue0.85mm}\hline\noalign{\vglue0.85mm}
$\chi_{c1}$ & \psip  &    2.61 $\pm$ 0.32 $\pm$ 0.08 \\
\noalign{\vglue0.85mm}\hline\noalign{\vglue0.85mm}
$\chi_{c2}$ & \psip  &    2.81 $\pm$ 0.34 $\pm$ 0.09 \\
\noalign{\vglue0.85mm}\hline\noalign{\vglue0.85mm}
$\Upsilon$(1S) & $\chi_{b0}$(1P) &  1.22 $\pm$ 0.24 $\pm$ 0.17 \\
                & $\chi_{b1}$(1P) &  21.7 $\pm$ 3.4 $\pm$ 1.1 \\
                & $\chi_{b2}$(1P) &  11.5 $\pm$ 2.0  $\pm$ 0.7 \\
                & $\Upsilon$(2S)      &  11.3 $\pm$ 1.5  $\pm$ 0.4\\
                & $\chi_{b0}$(2P) &  0.167 $\pm$ 0.033 $\pm$ 0.075 \\
                & $\chi_{b1}$(2P) &  5.15 $\pm$ 0.94 $\pm$ 0.48 \\
                & $\chi_{b2}$(2P) &  3.40 $\pm$ 0.63 $\pm$ 0.39 \\
                & $\Upsilon$(3S)      &  1.51 $\pm$ 0.28 $\pm$ 0.07 \\
                & $\chi_{b0}$(3P) &  0.018 $\pm$ 0.004 $\pm$ 0.016 \\
                & $\chi_{b1}$(3P) &  1.59 $\pm$ 0.32 $\pm$ 0.42 \\
                & $\chi_{b2}$(3P) &  1.35 $\pm$ 0.27  $\pm$ 0.45 \\
\noalign{\vglue0.85mm}\hline\noalign{\vglue0.85mm}
$\chi_{b0}$(1P) & $\Upsilon$(2S)     &   2.58 $\pm$ 0.54 $\pm$ 0.27 \\
                & $\Upsilon$(3S)     &   0.099 $\pm$ 0.024  $\pm$ 0.015 \\
\noalign{\vglue0.85mm}\hline\noalign{\vglue0.85mm}
$\chi_{b1}$(1P) & $\Upsilon$(2S)     &   4.73 $\pm$ 0.97 $\pm$ 0.29 \\
                & $\Upsilon$(3S)     &   0.033 $\pm$ 0.008 $\pm$ 0.019 \\
\noalign{\vglue0.85mm}\hline\noalign{\vglue0.85mm}
$\chi_{b2}$(1P) & $\Upsilon$(2S)     &   5.0 $\pm$ 1.0 $\pm$ 0.3 \\
                & $\Upsilon$(3S)     &   0.372 $\pm$ 0.087 $\pm$ 0.047 \\
\noalign{\vglue0.85mm}\hline\noalign{\vglue0.85mm}
$\Upsilon$(2S) & $\chi_{b0}$(2P) &  1.42 $\pm$ 0.30 $\pm$ 0.31 \\
                & $\chi_{b1}$(2P) &  19.0 $\pm$ 3.4 $\pm$ 1.7 \\
                & $\chi_{b2}$(2P) &  9.2 $\pm$ 1.8 $\pm$ 1.2 \\
                & $\Upsilon$(3S)  &  5.7 $\pm$ 1.1 $\pm$ 0.5 \\
                & $\chi_{b0}$(3P) &  0.15 $\pm$ 0.03 $\pm$ 0.11 \\
                & $\chi_{b1}$(3P) & 5.9 $\pm$ 1.2 $\pm$ 1.2 \\
                & $\chi_{b2}$(3P) & 3.7 $\pm$ 0.8 $\pm$ 1.0 \\
\noalign{\vglue0.85mm}\hline\noalign{\vglue0.85mm}
$\chi_{b0}$(2P) & $\Upsilon$(3S)   &     3.09 $\pm$ 0.73 $\pm$ 0.32 \\
\noalign{\vglue0.85mm}\hline\noalign{\vglue0.85mm}
$\chi_{b1}$(2P) & $\Upsilon$(3S)   &     6.5 $\pm$ 1.5 $\pm$ 0.6 \\
\noalign{\vglue0.85mm}\hline\noalign{\vglue0.85mm}
$\chi_{b2}$(2P) & $\Upsilon$(3S)   &     6.8 $\pm$ 1.5 $\pm$ 0.8 \\
\noalign{\vglue0.85mm}\hline\noalign{\vglue0.85mm}
$\Upsilon$(3S) & $\chi_{b0}$(3P) &  1.02 $\pm$ 0.25 $\pm$ 0.56 \\
                & $\chi_{b1}$(3P) &  17.0 $\pm$ 3.7 $\pm$ 2.5 \\
                & $\chi_{b2}$(3P) &  7.8 $\pm$ 1.8 $\pm$ 1.6 \\
\noalign{\vglue0.9mm}\hline
\end{tabular}
\end{table}

We have until now considered only \mbox{S-wave} states. 
Given the absence of detailed $\chi_b$ cross section data at mid rapidity, 
we can put to test the assumption that the ``universal'' $\sigma(\mathcal{Q})/\sigma(2\,m_Q) \propto E_{\mathrm{binding}}^{\delta}$ 
dependence of the long-distance factors, with $\delta \simeq 0.63$, can be extended to the \mbox{P-wave} states, 
even if with a different normalization constant multiplying $E_{\mathrm{binding}}^{\delta}$ 
to account for a dependence on the orbital angular momentum. 
In this way, $\chi_c$ data come to constrain the $\chi_b$(nP) cross sections, through the relation 
$\sigma(\chi_{bJ}{\rm (nP)}) = [ E_{\rm binding}(\chi_{bJ}{\rm (nP)}) /  E_{\rm binding}(\chi_{cJ}) ]^{\delta} \sigma(\chi_{cJ})$, 
corresponding to the assumption that there is no dependence on $J$, $n$, or quark flavour.

Using branching-ratio measurements from Ref.~\cite{bib:PDG}, 
the feed-down structure of quarkonium production can be fully predicted.
The result is presented in Table~\ref{tab:feeddowns} 
(see also Appendix~\ref{sec:appendix}).
It should be kept in mind that these results correspond to the $\pt/M > 2$ region.
The predicted feed-down fractions of $\Upsilon$(nS) production from $\chi_b$(nP) states 
are in reasonable agreement with forward-rapidity LHCb measurements 
merging the $J=1$ and $J=2$ signals~\cite{bib:LHCbChibFeedown}, considered for $\pt/M > 2$. 

\section{Summary and discussion}

At the current level of experimental precision, mid-rapidity LHC proton-proton data
for inclusive charmonium and bottomonium production are well described by a simple parametrization 
reflecting a universal (i.e.\ state-independent) scaling with two variables:
the shapes of the \pt distributions of all states become one common shape as a function of $\pt/M$ 
and the normalization of this shape scales in a simple monotonic way with $E_{\mathrm{binding}}$, 
at least for the \mbox{S-wave} states. 
While the cross section shape as a function of $\pt/M$ depends on $\sqrt{s}$, 
the normalization scaling with $E_{\mathrm{binding}}$ is found to be identical at 7 and 13\,TeV.

Having a simple empirical parametrization faithfully describing the proton-proton measurements 
is certainly very useful for model-independent studies of quarkonium production, 
especially when considering more complex collision systems such as proton-nucleus and 
nucleus-nucleus collisions. It is also very convenient for the tuning of Monte Carlo simulations.
More importantly, the analysis reported in this paper,
exclusively based on non-trivial (albeit potentially misleadingly simple) dimensional analysis arguments,
reveals significant experimental evidence supporting that quarkonium production can be
understood as being factorized between short- and long-distance phases. 
This data-driven result mirrors very well
the general concept of NRQCD factorization into process-dependent SDCs and universal LDMEs. 
More generally, seeing evidence of factorization in the LHC data is an important step towards establishing that
the \QQbar pairs are produced uncorrelated from the rest of the event,
allowing us to experimentally probe how they evolve and bind into the observable spectra of quarkonium states.

The experimental verification of the factorization ansatz can be extended 
by including in the analysis the available forward-rapidity LHCb measurements. 
Such study, currently ongoing, has a higher order of complexity: 
since the $\pt/M$ distributions show a significant $y$ dependence, 
the determination of scaling patterns is no longer an effectively one-dimensional problem 
as in the case where only mid-rapidity data are considered. 
Further tests, which may become possible in the future, 
include the study of the production of quarkonia associated to specific final state particles 
and the search for simplicity patterns in different kinds of collisions.

More data on the production of \mbox{P-wave} quarkonia are needed, especially in the bottomonium family:
the $\chi_b$ absolute cross sections have not yet been measured at mid rapidity, 
and the forward rapidity results, reported relative to $\Upsilon$(nS) production, 
do not distinguish between the $J = 0,1$, and 2 states.
It is reassuring, however, to see that those forward rapidity measurements are in reasonable agreement 
with the pattern of $\chi_{bJ}{\rm (nP)} \to \Upsilon$ feed-down fractions that we obtained 
assuming that the simple $E_{\mathrm{binding}}$ scaling pattern can be extended to the $\chi$ states,
an assumption that follows very naturally from the work presented in this paper.
The detailed feed-down structure of the charmonium and bottomonium families we have determined
in this work needs to be tested by new measurements, filling gaps in the experimental picture. 
%
It cannot be excluded that future measurements of direct $\chi_{cJ}$ and $\chi_{bJ}$ production 
will show that the scaling with binding energy is not the same for the P- and \mbox{S-wave} states. 
Such a result would still be very interesting, naturally. 
For example, the difference in scaling trends, when known, 
could be used to see if new resonances, as the X(3872), 
align better with the S- or the \mbox{P-wave} mass trends, 
thereby contributing to the understanding of their nature.

A complete account of all the feed-down fractions is particularly important in quantitative analyses 
of the quarkonium production and suppression patterns observed in nucleus-nucleus collisions.
It has been argued since long, in particular, that the observed suppression of the \jpsi\ signal, 
both from pp to heavy-ion collisions and from peripheral to central nuclear collisions, 
might be dominantly (or fully) caused by the suppression of the \psip\ and $\chi_c$ excited states,
more loosely bound and, hence, easier to dissolve in the hot medium created in those interactions.
Such hypotheses can only be reliably and quantitatively tested
if the hierarchy of feed-down relations between states is taken into account. 
This computation is now feasible, assuming the global $E_{\mathrm{binding}}$ scaling
that leads to the values collected in Table~\ref{tab:feeddowns} for pp collisions, 
complemented by a model for its modification in nucleus-nucleus collisions.
Such studies will finally address, and possibly provide evidence for, the concept of
sequential quarkonium suppression,
an ideal signature of quark-gluon plasma formation~\cite{bib:sequentialSuppression} that
applies to the directly-produced states, after the subtraction of the feed-down contributions.



\appendix
\section{Feed-down fractions from bottomonium to charmonium states.}
\label{sec:appendix}

Charmonium production through decays of bottomonia is usually assumed to be negligible. 
In Table~\ref{tab:feeddowns2} we provide quantitative constraints on the feed-down fractions
of some bottomonia-to-charmonia decays, 
calculated with the procedure described at the end of Section~3. 
The results confirm that, indeed, the two families can be considered as essentially disconnected.
 
\begin{table}[!htp]
\centering
\caption{Feed-down fractions between the $\Upsilon$(1S) and $\Upsilon$(2S) states 
and the several charmonia.}
\label{tab:feeddowns2}
\begin{tabular}{ccc}
\hline\noalign{\vglue0.85mm}
Daughter & Parent & Feed-down fraction (\%)\\ 
\noalign{\vglue0.85mm}\hline\noalign{\vglue0.85mm}
\jpsi          & $\Upsilon$(1S)     &   (5.57 $\pm$ 0.69) $10^{-5}$ \\
                & $\Upsilon$(2S)     &   (2.2  $\pm$ 2.2) $10^{-5}$ \\
\noalign{\vglue0.85mm}\hline\noalign{\vglue0.85mm}
$\chi_{c0}$ & $\Upsilon$(1S)    &    (3.4 $\pm$ 3.4) $10^{-5}$ \\
                & $\Upsilon$(2S)    &    (1.5 $\pm$ 1.5) $10^{-5}$ \\
\noalign{\vglue0.85mm}\hline\noalign{\vglue0.85mm}
$\chi_{c1}$ & $\Upsilon$(1S)    &    (4.26 $\pm$ 0.89) $10^{-5}$ \\
                & $\Upsilon$(2S)    &    (2.10 $\pm$ 0.55) $10^{-5}$ \\
\noalign{\vglue0.85mm}\hline\noalign{\vglue0.85mm}
$\chi_{c2}$ & $\Upsilon$(1S)    &    (7.1 $\pm$ 2.0) $10^{-5}$ \\
                & $\Upsilon$(2S)    &    (2.48 $\pm$ 0.92) $10^{-5}$ \\
\noalign{\vglue0.85mm}\hline\noalign{\vglue0.85mm}
\psip        & $\Upsilon$(1S)    &    (1.01 $\pm$ 0.22) $10^{-4}$ \\
                & $\Upsilon$(2S)    &    (0.35 $\pm$ 0.35) $10^{-4}$ \\
\noalign{\vglue0.9mm}\hline
\end{tabular}
\end{table}

\end{document}